\documentclass[aps,pra,showpacs,twoside,10pt,floatfix,nofootinbib,longbibliography,twocolumn]{revtex4-1}
\usepackage[colorlinks=true, citecolor=red, urlcolor=blue ]{hyperref}
\usepackage{epsfig,newlfont,amssymb,amsfonts,amsmath,bm,subfigure,palatino,mathtools,amsthm,braket,soul,enumitem,color,times,comment,geometry,float,array,physics}
\usepackage[table]{xcolor}
\usepackage[normalem]{ulem}
\begin{document}

\title{ Constructive impact of Wannier-Stark field on environment-boosted quantum batteries}
%\\Environment-assisted quantum battery using ultra-cold atoms with Wannier-Stark field}
\author{Animesh Ghosh$^{1}$, Tanoy Kanti Konar$^{1}$, Leela Ganesh Chandra Lakkaraju$^{1,2}$, Aditi Sen(De)$^{1}$}
\affiliation{$^1$ Harish-Chandra Research Institute, A CI of Homi Bhabha National Institute, Chhatnag Road, Jhunsi, Allahabad - 211019, India}
\affiliation{$^2$ Pitaevskii BEC Center and Department of Physics, University of Trento, Via Sommarive 14, I-38123 Trento, Italy }

\begin{abstract}

Using the ground states of the Bose- and Fermi-Hubbard model as the battery's initial state, we demonstrate that using the Wannier-Stark (WS) field for charging in addition to onsite interactions can increase the maximum power of the battery.  Although the benefit is not ubiquitous, bosonic batteries are more affected by the WS field than fermionic ones.  In particular, there exists a critical WS field strength above which the power gets increased in the battery. Further, we determine a closed form expression of the stored work when the battery is in the ground state of the Bose- and Fermi-Hubbard model with only hopping term and the charging is carried out with onsite interactions and WS field  irrespective of lattice-size of the battery. Moreover, we exhibit that it is possible to extract work in the fermionic batteries even without  charging when  the edge sites are attached to two local thermal baths having high temperatures -- this process we refer to as  {\it environment-assisted ergotropy}. Note, however, that the bosonic batteries are able to exhibit such an environmental benefit in the transient regime when the lattice-size is increased and when Wannier-Stark field is present.   Nonetheless, if the onsite interaction or WS potential with a critical strength is utilized as a charger, energy can be stored and extracted from both bosonic and fermionic batteries in the presence of the thermal baths.  

\end{abstract}
\maketitle

\section{Introduction}

%Battery is a device in which we can store energy and also extract energy from it whenever we wish. Here we are trying to build battery using principles of quantum mechanics \cite{konar_pra_battery_2022}.

Better storage techniques are more important than ever as the world's energy consumption rises daily. To address this,
quantum batteries (QBs) have been proposed \cite{Alicki, battery_rmp_review}, where the battery is charged through unitary operations after being prepared in either a product or an entangled state \cite{campaioli2017,LewensteinLR,Modispinchain, srijon2020, srijon2021}.  It has also been demonstrated that quantum batteries, which leverage aspects such as entanglement \cite{RMPentanglement}, coherence \cite{RMPcoherence}, and the entangling capabilities of unitaries \cite{campaioli2017}, outperform classical batteries that lack these quantum features in terms of energy storage and extraction capacity.  Numerous theoretical developments \cite{andolina2017,andolina2019,santos2019,andolina2020, Bera2020QB,ksen_battery_1,alba_1_20,alba_1_22,floquet_battery,konar_battery_2,konar_battery_3,ksen_battery_4,arjmandi_pra_2022,santos_pra_2023,ksen_battery_5,AI_quantum_battery,sashi_pra_1,mesure_battery,remote_charging_battery,topological_quantumbattery,NV_battery,sashi_2,perciavalle2024,stavya_battery} and experimental demonstrations in various physical systems~\cite{wennigerexpqdots,superconducting_battrey_1, superconductQBexp, GemmeexpIBMsupercond, Quach2022Jan, MaheshexpNMR} have been carried out since its inception. These include  various protocols for charging and discharging \cite{remote_charging_battery}, such as chain-stimulated Raman adiabatic passage~\cite{stirap_battery_2}, measurement-based schemes~\cite{mesure_battery}, the use of geodesic paths to minimize charging~\cite{Halder2024Mar}, indefinite causal order~\cite{battery_ico}, etc., and the utilization of many-body quantum phenomena, such as localization~\cite{arjmandi_pre_2023}, to improve the performance  of the QB.

Ultracold atomic systems, alongside other physical platforms, offer an excellent platform for investigating basic quantum many-body physics  \cite{bloch03, aditi07,BlochRMP08} and constructing quantum technologies \cite{AmicoRMP22, QmetroRMP} including quantum computers and quantum batteries. The ability to ``engineer" quantum systems in a controllable manner makes cold atomic platforms particularly promising for future quantum technologies \cite{BlochRMP08, Bloch12}.  In this model, the competition between kinetic energy (the hopping of atoms between lattice sites) and interaction energy (onsite repulsion) leads to two distinct quantum phases: the Mott-insulator phase, which results in a highly localized state because atoms are prevented from hopping and the superfluid phase \cite{bloch03}, which allows atoms to move freely between sites. One landmark achievement in this field was the first experimental observation of a quantum phase transition, realized in the Bose-Hubbard model.  Consequently, these systems have also proven  to be an intriguing system when analyzed within the framework of quantum thermodynamics, which allows for the investigation of issues pertaining to energy extraction, storage, and transfer in quantum systems \cite{konar_pra_battery_2022}.

The delocalization effect in the superfluid phase can be tailored by introducing a static field into the system, specifically the Wannier-Stark electric field \cite{Stark1913Dec,Wannier1960Apr,wannier_pr_1960}. Specifically, it was shown that the strength of the Wannier-Stark field can cause changes in the energy spectrum of the Hubbard models \cite{WStarkKrishnendu02,WStarkBuchleitner03,WStarkKolovsky03}. Notably,   the ground state of this model for a single particle is extremely localized, known as Stark localization, and can exhibit several exotic phenomena both in both fermionic and bosonic models \cite{krimer_pre_2009,haque_prl_2020,gao_prb_2023,mina_prb_2023,wohlman_prb_2024}. The alteration in spectral characteristics caused by the static field naturally raises the question of how the Wannier-Stark ladder influences the energy storage capability of QB in both Bose-Hubbard (BH) and Fermi Hubbard (FH) models under both unitary and dissipative dynamics, which is the main focus of this work. 

Starting from the ground states of the BH and FH models with nonvanishing hopping and onsite interaction strengths, we derive analytical expressions of the work-output of a battery  with in the presence of onsite interactions and Wannier-Stark potential in the charger  which clearly illustrates the beneficial role of charging parameters. These results highlights the beneficial role of charging parameters, showing that the maximum average power increases with the static field strength in the charging Hamiltonian beyond a certain threshold.  However, the WS field in the battery itself exerts a destructive influence on power generation. For moderate  onsite interaction strength, bosonic batteries was shown to generate less power than the fermionic ones in the absence of an electric field  \cite{konar_battery_1}.  Interestingly, introducing a static field in the charging Hamiltonian reverses this trend, a phenomenon we term {\it activation of power} with the aid of the WS field. This activation is more pronounced when the initial state is prepared at moderately high temperatures compared to low temperatures. We further show that the WS field during charging enables power to scale with the number of lattice sites, albeit sublinearly, whereas without the field, power declines with increasing lattice size.  On the other hand, the increment in the particle numbers of QB having WS field as a charger leads to the saturation of power except a very high value obtained for a single particle which is much higher than the one obtained without the static field. These findings emphasize the crucial role of the WS field in enhancing power generation and storage in quantum batteries, offering insights into optimizing performance under both bosonic and fermionic setups.

Even though some of us have suggested an ultracold atomic quantum battery \cite{konar_battery_1, stirap_battery_2}, their behavior under environmental influence, such as lattice sites connected to a bath, remains unexplored. Previous studies have shown that baths can induce long-range correlations \cite{tindall_prl_2019,ribeiro2023,saha2024}, and exotic phenomena in quantum systems  \cite{eckdart_1,eckdart_5,Kordas2015Nov,eckdart_4,eckdart_2,eckdart_3,schaller_prb_2022,christian_prr_2023,Marco_prb_2024,marco_review_2024}. In this paper, we show that  extractable work (i.e., the nonvanishing ergotropy)  can be stored in the QB whose  edge sites  are attached to the thermal bath having same or different temperatures.  Remarkably, in the presence of hopping, onsite interactions, and a Wannier-Stark (WS) field, fermions exhibit steady-state ergotropy even without a charger, while bosons achieve this only in a transient domain as the system size increases -- we refer to this phenomenon as {\it environment-assisted ergotropy}.   Unlike closed dynamics, Wannier-Stark ladder does play a favorable  role for bosons and fermions to discharge the battery with the help of environment. Furthermore, we observe a non-monotonic behavior of ergotropy in the steady state for bosons as well as fermions  with the increase in the strength of the WS field and the onsite interactions during charging, underscoring the advantage of moderate WS field  in QB design.

The paper is organized in the following manner. Sec. \ref{sec:dependencyStark} introduces the quantum battery set-up and presents the analytical results for the work-output when  Wannier-Stark potential is used for charging (closed dynamics). The activation of power with the help of WS field for bosonic battery over fermionic ones is reported in Sec. \ref{sec:activation} which also includes the scaling analysis of QB with nearest-neighbor  interactions. The set-up in which the extractable work is stored with the help of the thermal bath attached to the edges of the chain is discussed  in Sec. \ref{sec:noiseB}. The concluding remarks  is included in Sec. \ref{sec:conclu}.

\section{Dependency of power upon Wannier-Stark strength} 
\label{sec:dependencyStark}

Batteries made up of ultracold atoms in an optical lattices have already been designed \cite{konar_battery_1} and  was shown that the power extracted from fermionic batteries are, most of the situations, higher than that obtained from the bosonic ones. By incorporating the static electric field in the Hubbard Hamiltonian, our aim is to exhibit whether any benefit in the storage capacity of energy in quantum batteries, especially, in bosonic case, can be obtained due to the emergence of Wannier-Stark  ladder in the energy domain \cite{wannierreview} and whether it can provide more robustness against decoherence. 

{\it Battery set-up.} Let us prepare the QB in the ground state of the Bose-Hubbard model with Stark energy \cite{wannierreview}, given by
\begin{eqnarray}
     H_{B} &=& -J_B\sum_{\langle i,j\rangle}   \hat{b}^\dagger_i \hat{b}_{j}  +  \text{h.c.}  +  \sum_{i} \frac{U_B}{2} \hat{n}_i (\hat{n}_i-1) \nonumber \\
     &-&r_B\sum_{i} i \hat{n}_i,
    \label{eq:BHbattery}
\end{eqnarray}
where \(\hat{b}_i^\dagger\) (\(\hat{b}_i\)) is the creation (annihilation) operator of one boson at site \(i\) which follows the bosonic commutation rule, given as \([\hat{b}_i,\hat{b}_j^\dagger]=\delta_{ij}\), \(\hat{n}_i=\hat{b}_i^\dagger \hat{b}_i\), \(J_B\) is the strength of hopping between nearest-neighbor lattice sites, \(i\) and \(j= i+1\), \(U_B\) is the onsite interaction strength and \(r_B\) defines the Wannier-Stark interaction strength in this model. \(N\) denotes the lattice-size and \(n= =\sum_i\langle n_i\rangle\) represents the number of particles. 
% We consider here two kinds of battery Hamiltonian  -- (1) \(J \neq 0\), (2) \(U\, \text{and/or}\, r_B \neq 0\). In the former case, the charger contains BH model with \(U^c\, \text{and/or}\, r_B^c \neq 0, \) while in the latter case, \(J^c \neq 0\). 
% 

When bosonic operators are replaced by the femionic ones, we obtain the Fermi-Hubbard model in presence of static field as  
\begin{eqnarray}
     H_{F} &=& -J_F\sum_{\langle i,j\rangle, \sigma}  \hat{c}^\dagger_{i \sigma} \hat{c}_{j \sigma}  +  h.c.  +  \sum_{i} U_F \hat{n}_{i \uparrow}\, \hat{n}_{i \downarrow}  \nonumber \\
     &-& r_F\sum_{i} i (\hat{n}_{i\uparrow}+\hat{n}_{i\downarrow}),
    \label{eq:FH}
\end{eqnarray}
where \(\hat{c}^\dagger_{i \sigma} (\hat{c}_{i \sigma})\) is creation (annihilation) operator of \(\sigma \in \{\uparrow,\downarrow\}\) spins at site \(i\) which follows the fermionic anti-commutation rule. \((\hat{n}_{i\uparrow}, \hat{n}_{i\downarrow})\) is the number operator at site \(i\) corresponding to \(\uparrow\) and \(\downarrow\) spins respectively. 
The main aim of this work is to identify the role of \(r_B\) or \(r_F\) in the 
%power {\color{red} work output (neither power nor work are not defined)} 
performance of the QB.  
%{\color{red} Note that superscripts are used for the charging  Hamiltonian and subscripts describe the statistics that the model is following while no subscripts are used when the same value is used both for bosons and fermions.} 

%By imposing the particles per site upto two and including the Wannier-Stark potential, let us first derive the analytical expression for the work output where the number of particles is taken to be same as the lattice sites, i.e., filling factor, \(\expval{n}/N=1\).

{\bf Proposition 1.} {\it If the battery with two sites  is prepared in the ground state of the BH and FH Hamiltonian  having only  hopping term,  the work-output in both the cases becomes \( (1-\cos(r^c_x t) \cos(U^c_x t))\, (x=B, F)\) where the battery is charged via onsite interaction and  Wannier-Stark potential.}
% where \(A'\) is functions of the hopping, Wannier-Stark and onsite interaction strength. 

\begin{proof}
%\adi{one has to write the initial state and evolving Hamiltonian and final state and work output. }
The Bose-Hubbard model with N sites and n particles has the set of basis states,  {$\{\ket{i_1, i_2, ... i_N}\}$ such that $\sum_\mu i_\mu = n$ and $i_\mu \leq 2 \,\, \forall \mu \in [1,N]$} and for the Fermi-Hubbard model, the basis is defined as $|x_{\uparrow} y_{\downarrow}\rangle_{1} |z_{\uparrow} w_{\downarrow}\rangle_{2}\ldots$ where
\(\{x, y, z, w\} \in (0,1)\), where \(0\) denotes the situation when the lattice site is not occupied by a fermions while \(1\) is when the fermion occupies the lattice site, \(i\), subscripts denote the lattice sites which we drop now onward and  we will use only the binary method to indicate the entire configuration. The initial states of the battery Hamiltonian with \(N=2\) having only hopping term for bosons and fermions at zero temperature read  as 
%$H^B_b = \left(
% \begin{array}{ccc}
%  0 & -\sqrt{2} J & 0 \\
%  -\sqrt{2} J & 0 & -\sqrt{2} J \\
%  0 & -\sqrt{2} J & 0 \\
% \end{array}
% \right)$, 
$\ket{\Psi^{in}_B} = \frac{1}{2} \ket{0,2} + \frac{1}{\sqrt{2}}\ket{1,1} + \frac{1}{2}\ket{2,0}$ and $\ket{\Psi^{in}_F} = \frac{1}{2} (\ket{0,0,1,1} + \ket{0,1,1,0} + \ket{1,0,0,1} + \ket{1,1,0,0}$)
respectively. Here odd and even places denote up spins and  down spins respectively for fermions. For bosons, the charging Hamiltonian with onsite and Wannier-Stark potential takes the form,
$H^c_B = (U^c_B -4 r^c_B) \ket{0,2}\bra{0,2} + (-3 r^c_B) \ket{1,1}\bra{1,1} + (U^c_B-2 r^c_B) \ket{2,0}\bra{2,0}$,
% $H^c_B = \left(
% \begin{array}{ccc}
%  \text{U}^c_B -4 \text{r}^c_B & 0 & 0 \\
%  0 & -3 \text{r}^c_B & 0 \\
%  0 & 0 & \text{U}^c_B-2 \text{r}^c_B \\
% \end{array}
% \right), $ 
resulting in the final state $\rho^{fin}_B $, for arbitrary time \(t\) and, similarly, for fermions. Note that superscripts in the parameters are used for the charging  Hamiltonian and subscripts describe the statistics of the model while no subscripts are used when the same value is used both for bosons and fermions.
In both the situations, the  expression for the work output is given as
\begin{eqnarray}
    W_x(t) =\Tr [H_x(\hat{\rho}^{fin}_x-\hat{\rho}^{in}_x)] =  1 - \cos (r^c_x t) \cos(U^c_x t) \,
\end{eqnarray}  
with \(x = B, F\). Note that to make a fair comparison, we have to normalize the Hamiltonian as\\ 
\( \frac{1}{E_{\max}-E_{\min}}[2H_B-(E_{\max}-E_{\min})\mathbb{I}]\rightarrow H_B 
\)
 and hence \(W_x(t)\) becomes independent of the strength of hopping and is a function of parameters involved in the charging.
\end{proof}

%First, consider a scenario in which two particles occupy two sites. In this situation, the work output can be found analytically for both bosons and fermions. We choose the battery Hamiltonian is made of hopping term, \(J\) and onsite interaction strength, \(U\) and we charge the Hamiltonian with Wannier-Stark potential,\(r^c\) and the onsite interaction strength, \(U^c\).  The expression of work output is given as
%$ 1 - \cos (r^c t) \cos(U^c t)$
% \begin{equation}
%     W(t)=\frac{16 J^2 }{\sqrt{16 J^2+(U^{c})^2}} [1-\cos (r^c \, t) \cos (U^c \,t)].
% \end{equation}
%\adi{this above equation considers battery Hamiltonian to be J, so denomenator mein \(U^c\) hoga na? I changed}

One can observe that the stored work is same for both FH and BH models although  they obey different statistics which is due to the particular design of the battery, the charger and two sites. Further, \(W_x(t)\) at a particular time is a periodic function of \(r^c_x\) and  \(U^c_x\). It implies that certain values of Wannier-Stark and onsite interaction strength are optimum in order to store the maximum work, i.e.,  \(r^c_x=U^c_x=\frac{ p \pi}{t}\) (\(p \in Z\)) which is independent of other battery parameters, like the signs of $r^c_x$ and $U^c_x$. If both $r^c_x$ and $U^c_x$ are present, the work will be periodic with period = LCM[$\frac{2\pi}{r^c_x}$ and $\frac{2\pi}{U^c_x}$]. Therefore, the maximum value of the work can be obtained at the time which is half of this time period. 

Let us now consider the average power, which is given as $\frac{W_x(t)}{t}$, and the maximum average power, defined as 
\begin{eqnarray}
    P^{\max}_x = \underset{t}{\max} \,\frac{W_x(t)}{t},\, \, x=B,F,
    \label{eq:maxpower}
\end{eqnarray}
where the maximization is performed over the time of the charging.  
 Note, however, that \(P^{\max}_x\) is not maximum around $t=\frac{ p \pi}{r^c_x}$ since it depends upon several other parameters, including time and charging Hamiltonian. This situation depicts that the maximization of both work and power does not fall in the same time which describes that achieving faster charging speed does not necessarily imply to have maximum work storage.

% It is natural to ask that among these two set-up, which one is the best battery Hamiltonian and charger for bosons? The condition  between system parameters dictate the ordering between their work-outputs. \\

%{\it Proposition 3. } \adi{Conditions to be put and found in which Proposition 1 is better than Proposition 2. }

% For Bose-Hubbard model, we have analytically calculated the expressoin of work when battery Hamiltonian is the standard Hubbard model(U+J) and charging Hamiltonian consists of U and $r_c$ terms. The expression looks like:
% \begin{equation}
%     W(t)=\frac{32 j^2 (\cos (r_c t) \cos (U_c t)-1)}{U \left(\sqrt{16 j^2+U^2}+U\right)+16 j^2}
% \end{equation}

{\it Quantum Battery with arbitrary lattice-size: bosons and fermions.} 
Let us consider the same QB set-up as described in Proposition 1 with arbitrary lattice-size having unit filling factor. Although the work-output and the maximum average power in this situation cannot be obtained analytically,  the careful numerical simulations suggest the expressions for the work-output for both bosons and fermions as 
% Now, we generalize the result for arbitrary lattice-size where filling factor is one. Although, the exact generalization is not possible, we provide a qualitative expression of work stored as well as average power which is given as
\begin{equation}
    W_x^N(t)=\alpha + \beta \cos(r^c_x\,t) + \gamma \cos(r^c_x\,t) \cos(U^c_x\,t),\, x= B, F.
    \label{eq:generalworkwithWS}
\end{equation}
Here \(\alpha, \beta\), and \(\gamma\) are functions of the  parameters in the charging Hamiltonian, \(J^c_x,U^c_x\), and \(r^c_x\) which also depend on the statistics of the particles. 
Therefore, to study the performance of the battery, we require the explicit forms of $\alpha, \beta$ and $\gamma$ which we will use in the succeeding section for comparing the bosonic and fermionic batteries. Before proceeding further, let us concentrate on the scenario in which the battery Hamiltonian consists of either Wannier-Stark potential or onsite interaction while the charging is performed by allowing hopping between the particles. In this case, although bosonic and fermionic batteries behave similarly in the former case as also in Proposition 1 for two lattice sites, bosons always outperform fermionic batteries in the latter case,. 
% Similar to the two-site lattice, work stored in a \(N\) site lattice is also independent of the sign of the Wannier-Stark potential. In order to compare FH battery with BH battery we plot the difference of stored work in a particular time with variation of charging strength \(r^c\) and \(U^c\) in Fig. which shows that if \(r^c=0\) FH battery store more wotk than the BH battery. On the other hand, by the introduction of Wannier-Stark potential BH battery beats BH battery which show that contrast behavior.

%\adi{PROOF HAS TO BE CHANGED.}
%{\it Two-site lattice: boson vs. fermion} 

{\bf Proposition 2.} 
{\it If the battery with two lattice sites is prepared as the ground state of the Bose- and Fermi-Hubbard models with only onsite interaction, the stored work by bosons is always double of that of the fermions provided the particles are allowed to hop during the charging while bosons and fermions provide the same work-output when the initial state is in the ground state of the Wannier-Stark potential.}
% If the initial state of the battery  is the ground state of the battery Hamiltonian containing  either on-site  or Wannier-Stark interaction, and the system evolves according to \(H_{hop}\) where \(H_{hop}\) involves the hopping term of the Bose-Hubbard Hamiltonian with two modes, the work-output of the resulting state  is periodic with time and reads as \(B \sin^2(At)\) where \(A\) and \(B\) are functions of the system parameters. 
\begin{proof}
% \adi{one has to write the initial state and evolving Hamiltonian and final state and work output. }
For bosons, the initial state is $
% H_B (J_B = r_B = 0)$ reads as  
% \\$H^B = \left(
% \begin{array}{ccc}
%  U & 0 & 0 \\
%  0 & 0 & 0 \\
%  0 & 0 & U \\
% \end{array}
% \right), 
\,\rho_B^{in}=\ket{1,1}\bra{1,1}$
% \left(
% \begin{array}{ccc}
%  0 & 0 & 0 \\
%  0 & 1 & 0 \\
%  0 & 0 & 0 \\
% \end{array}
% \right)$
and the corresponding charging Hamiltonian is given by 
$H^c_B = -\sqrt{2} J^c_B \,\, (\ket{0,2}\bra{1,1} + \ket{1,1}\bra{0,2} + \ket{2,0}\bra{1,1} + \ket{1,1}\bra{2,0})$
% \left(
% \begin{array}{ccc}
%  0 & -\sqrt{2} \text{J}^c_B & 0 \\
%  -\sqrt{2} \text{J}^c_B & 0 & -\sqrt{2} \text{J}^c_B \\
%  0 & -\sqrt{2} \text{J}^c_B & 0 \\
% \end{array}
% \right)$
\\leading to the work output = $2 \sin^2 (2 J^c_B t)$.

On the other hand, in case of fermions,
denoting \(\{\ket{1,1,0,0}, \ket{1,0,0,1}, \ket{0,1,1,0}, \ket{0,0,1,1}\}\) as \(\{\ket{0}, \ket{1}, \ket{2}, \ket{3}\}\), 
% \\$H^B_f = \left(
% \begin{array}{cccc}
%  \text{U}^B_F & 0 & 0 & 0 \\
%  0 & 0 & 0 & 0 \\
%  0 & 0 & 0 & 0 \\
%  0 & 0 & 0 & \text{U}^B_f \\
% \end{array}
% \right),
we take 
$\rho_F^{in} = \frac{1}{2}\,\,(\ket{1}\bra{1} + \ket{2}\bra{2})$ as the initial state
% \left(
% \begin{array}{cccc}
%  0 & 0 & 0 & 0 \\
%  0 & 0.5 & 0 & 0 \\
%  0 & 0 & 0.5 & 0 \\
%  0 & 0 & 0 & 0 \\
% \end{array}
% \right)
and 
$H^c_F = -\text{J}^c_F\,\, (\ket{0}\bra{1} + \ket{1}\bra{0} + \ket{0}\bra{2} + \ket{2}\bra{0} + \ket{3}\bra{1} + \ket{1}\bra{3} + \ket{3}\bra{2} + \ket{2}\bra{3})$,
which results in $W_F(t) = \sin^2(2 J^c_F t)$. It implies $W_B(t) = 2 W_F(t)$, i.e.,   the work output obtained from bosons  is double to that of fermions.

If the battery Hamiltonian $H_B$ only contains  $r_B \neq 0$ with $J_B = U_B = 0$ and similarly, for fermions, $H_F$ having $r_F \neq 0$ ($J_F = U_F = 0$), the work-output for both bosons and fermions takes the form $2\sin^2(J^c_x \, t) (x = B,F)$, where the charging is carried out with the BH and FH models having only hopping term. 

\end{proof}

\begin{figure}
    \centering
    \includegraphics[width=\linewidth]{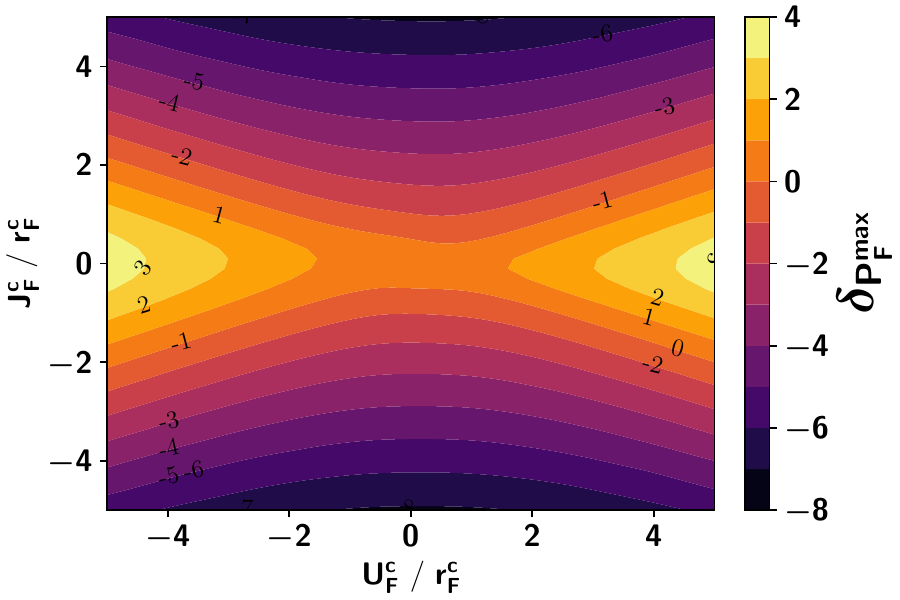}
    \caption{Difference between maximum powers, \(\delta_{P^{\max}_F}\),  obtained from two battery Hamiltonian -- (C1) one contains hopping term; (C2) it contains only Wannier-Stark term in the Fermi-Hubbard model. The charging is performed by changing the ratio of  onsite interaction and Wannier-Stark field, \(U^c_F/r^c_F\) , (horizontal axis)  in the case of C1 and by varying the ratio  of  \(J^c_F/U^c_F\) (vertical axis)  for the  case, C2. The initial state is prepared as the ground state of the battery Hamiltonian. We have taken fermions with two up and two down spins  in four sites. The similar results can be obtained for bosons with four site and four bosons.  Both the axes are dimensionless. }
    \label{fig:roleWSQB}
\end{figure}

To visualize the role of Wannier-Stark term on the performance, both in the battery  and the charging Hamiltonian, we compute the maximum average power by  considering two situations  -- Case 1 (C1). \(J_x \neq 0,\,r_x=U_x =0\) in the battery Hamiltonian with the variation of \(U^c_x\), and \(r^c_x\) during charging  and Case 2 (C2).  \(r_x \neq 0,  \, U_x = J_x =0\), by varying   \(J^c_x\), and \(r^c_x\) in the charging Hamiltonian (\(x=, B, F\)). To identify the regimes where the maximum power can be achieved, we are interested to calculate 
\begin{equation}
    \delta_{P^{\max}_x} = P_x^{\max} (C1) - P_x^{\max} (C2).
    \label{eq:deltaP}
\end{equation}
 Also we have constrained the number of bosons in a single site to be maximum two which is true for fermions due to Pauli exclusion principle. Firstly, we note that \(\delta_{P^{\max}_x} \)  is independent of the nature of particles, i.e., bosons and fermions although we find fermions can produce more power than that of bosons, i.e.,  \(P^{\max}_F \geq P^{\max}_B\) without the WS field which no more remains true in the presence of WS field.  Secondly, we observe that the difference gets maximized when \(J^c_x\) vanishes since the battery contains the hopping term and hence the hopping in the charging does not play any role. Thirdly,  the minimum is attained, i.e.,  Case 2 is favorable over  Case 1 when the strength of hopping, \(J^c_x \), is high  irrespective of the values of 
\(U^c_x\) and \(r^c_x\) (see Fig. \ref{fig:roleWSQB}). It is evident from this analysis that if the battery Hamiltonian contains hopping, the charging Hamiltonian only requires onsite and Wannier Stark potential to generate the power and vice-versa. In the rest of the paper, we will concentrate on Case 1 and vary other system and charging parameters to store or extract energy from the battery.

{\it Single particle present in QB with WS field as a charger.} Let us discuss the work stored and the power with a single particle in the lattice. As one can observe, due to a single particle, onsite interaction and particle statistics do not have any role to play. Hence we explicitely discuss the situation for bosons. 

If the battery is made up of hopping term, i.e., \(H_B=-J_B\sum_{i}\hat{b}_i^\dagger \hat{b}_{i+1} + h.c.\) and  the charging of the QB is performed only through the Wannier-Stark potential, the model can be solved exactly, and, therefore, the work stored in the battery can be obtained in the thermodynamic limit, i.e., \(N \rightarrow \infty\). In particular,  the initial state is taken to be the ground state of the QB, given as \(\ket{\Psi(t=0)}=\sqrt{\frac{2}{N+1}}\sum_{j=1}^{N}\sin\left (\frac{j\pi}{N+1}\right )\ket{j}\), where \(\ket{j}\) is the computational basis and the corresponding eigenvalue is \(E_{gs}=-2J_B\cos\left (\frac{\pi}{N+1}\right)\). By evolving the state with  the Wannier-Stark interaction, the evolved state takes the form as \(\ket{\Psi(t)}=\sqrt{\frac{2}{N+1}}\sum_{j=1}^{N}e^{-ijr^c_B t}\sin\left (\frac{j\pi}{N+1}\right )\ket{j}\), leading to the work stored  in the QB, given by 
\begin{eqnarray}
    W_B(t)= 2 J_B\cos\left(\frac{\pi}{N+1}\right ) (1-\cos (r^c_B t)).
    \label{eq:singleparticlebosons}
\end{eqnarray} 
As \(N\to\infty\), the work stored in the battery saturates to 
\(2J_B(1-\cos(r^c_Bt))\). It also indicates that the average power scales as \(\sim \cos\left(\frac{\pi}{N+1}\right )\).
%which is much lower than the classical scaling \(\sim N\), i.e., QB made of  a single particle does not provide any advantage in scaling, i.e., with the increment of lattice-size. This analysis leads to the following Proposition. 

{\bf Proposition 3.} {\it For bosonic and fermionic batteries having only hopping \(J_x \neq 0\) term which is charged  by the Wannier-Stark field, i.e.,  with nonvanishing \(r^c_x\) and  \(U^c_x =0\), the work stored (after normalization) for any arbitrary sites and arbitrary particle number becomes 
\begin{eqnarray}
    W_x(t) = 1 - \cos (r^c_x t) \,\, (x= B, F). 
\end{eqnarray}}
The above proposition immediately indicates that the impact of particle numbers and lattice-size on the performance of the quantum battery may play a  non-trivial role, when \(J_x\), \(U_x\), \(r_x\) of the battery, \(J^c_x\),  \(U^c_x\), and \(r^c_x\) are all non-vanishing. 

% Both the batteries are half filled (number of sites = number of particles), and the bosonic battery is constrained to have at most 2 particles in one single site. Since we are considering closed system here, the time-evolution will be unitary
% Work stored in this battery has a general form: $a + b \cos(r_c\,t) + c \cos(r_c\,t) \cos(U_c\,t).\,$ Here a,b,c are functions of J and U. b and c are negative.
% We can see, work is same for both positive $r_c $ and negative $r_c$ as cosine function is symmetric with respect to $r_c$. If $r_c = 0$, that means we charge the battery using only $U_c$, then stored energy in fermionic battery is always greater than its bosonic counterpart for $n > 2.$ But when we add the Wannier-Stark potential term in the charging Hamiltonian, Bosonic battery always stores more energy than the fermionic one.

% \begin{equation}
%     P(t) = \frac{W(t)}{t} = \frac{a+b \cos(r_c t)+c \cos(r_c t) \cos(U_c t)}{t}
% \end{equation}

\section{Activation of power in bosonic battery via Wannier-Stark potential over fermionic battery}
\label{sec:activation}

 In the entire discussion, hopping is always included in the battery Hamiltonian, the other parameters, onsite and Wannier-Stark ladder, are present both during  charging and in the initial state preparation of the battery.   Note first that  with the introduction of Wannier-Stark ladder in the battery Hamiltonian, the power decreases as \(r_x\) increases, thereby showing detrimental effect in QB under unitary evolution. 
 
\begin{figure}[h]
    \centering
    \includegraphics[width=0.8\linewidth]{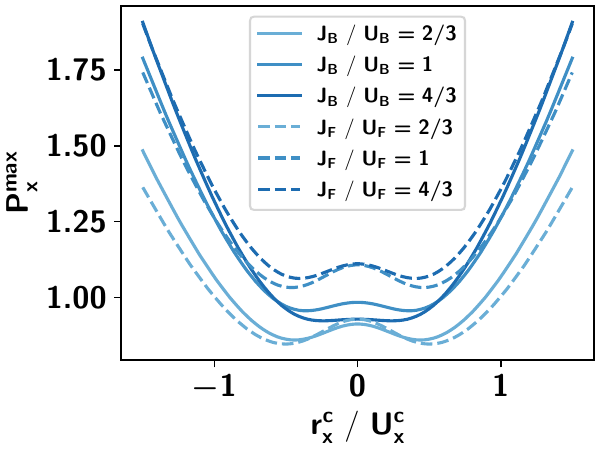}
    \caption{Variation of maximum power, \(P_{x}^{\max}\) (\(x= B, F\)) (ordinate) with respect to the strength of Wannier-Stark term (\(r^c_x/U^c_x\))  (abscissa) of the charging Hamiltonian. The subscripts are not given as curves represent behaviors of both bosons and fermions.  Light to dark shade curves indicate the increase of the ratio between  the hopping and the onsite interaction strength in the battery Hamiltonian. Solid and dashed lines  represent bosons and fermions respectively. All of them is computed for four particles with four lattice-size. Clearly, power behaves nonmonotonically with the strength of the WS field.  Both the axes are dimensionless. }
    \label{fig:roleofrc}
\end{figure}

 {\it Emerging nonmonotonicity in power with Wannier-Stark field in charging. }  If WS field is incorporated in the charging Hamiltonian, \(P^{\max}_x\) increases with \(r^c_x\) after a critical \(r^c_x\) strength.  Interestingly, however, some nonmonotonic behavior emerges. In particular, when $0\leq |r^c_x| < |r'^c_x|$, the power decreases with the increase of \(|r^c_x|\) while it monotonically increases with \(|r^c_x|\) beyond the threshold value, \(|r'^c_x|\) (Fig. \ref{fig:roleofrc}). However, this nonmonotonic nature of power can be eradicated by increasing the ratio between the hopping  and onsite interaction strength  in the  battery Hamiltonian for bosons although such an elimination  is not possible for fermions. Looking at the increment of power in QB with \(|r^c_x|\), we involve moderate amount of Wannier-Stark field strength during the charging in the entire paper.  

% \begin{figure}[h]
% \includegraphics[width=0.45\textwidth]
% {6bosonsU3_Jc0_Uc_2.pdf}
% % {J=1,r=0,U=3,r vary,U_c=2,Pmax,BosonsVsFermions.pdf}
% \caption{}
% % {Wannier-Stark term in battery Hamiltonian is not helpful. We have taken J=1,r varying, U=3, $r_c=0$, $U_c=2$}
% \label{}
% \end{figure} 

% \begin{figure}[h]
% \includegraphics[width=0.4\textwidth]
% {J=1,r=0,U=3,r_c vary,U_c=2,BosonsVsFermions.pdf}
% \caption{Charging becomes quicker if we introduce wannier-stark ladder in charging Hamiltonian. We have taken J=1, r=0, U=3, $U_c=2$}
% \label{}
% \end{figure}

\subsection*{Activation of power in bosonic battery over fermionic one} 

\begin{figure}
    \centering
    \includegraphics[width=\linewidth]
    {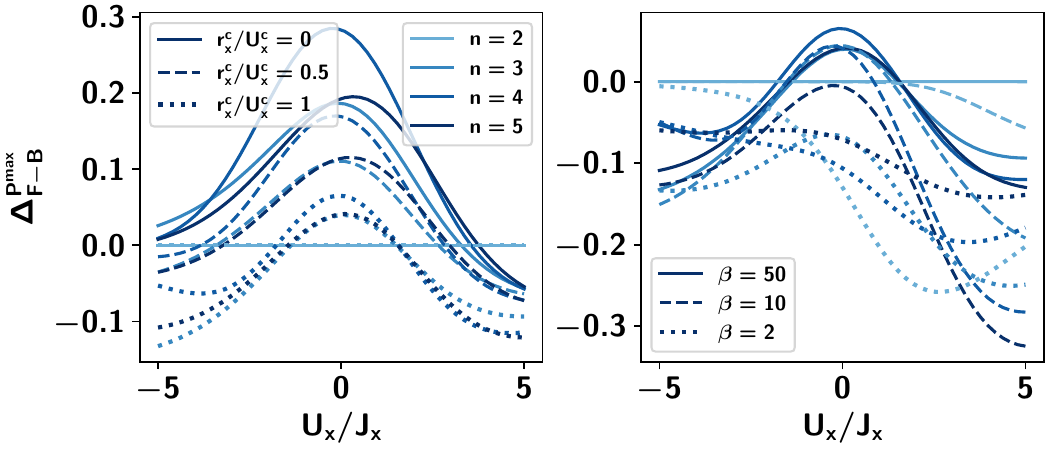}
    \caption{ Variation of maximum power, $\Delta_{F-B}^{P^{max}}$ (ordinate) with respect to the strength of onsite interaction, \(U_x/J_x\))  (abscissa) of the battery Hamiltonian.  In (a) \(\beta =100\) for different values of \(r^c_x/U^c_x\) and in (b), we choose different \(\beta\) values and  fix \(r^c_x/U^c_x =1\).  Here \(n= N\) for bosons and \(N/2= n_{\uparrow} = n_{\downarrow}\) for fermions with even \(N\) while  \(N+1/2= n_{\uparrow},  (N-1)/2= n_{\downarrow}\).  The more it gets negative, bosons tend to perform better. Solid, dashed and dotted curves represent different strengths of WS term in the charging Hamiltonian. Increment of shades of the plots represents increment of particle number. Note that when there are two particles and \(r^c =0\), bosons and fermions can generate same amount of power while in the presence of nonvanishing \(r^c_x/U^c_x\), bosons start to produce more power than the corresponding fermionic battery.   All of them is computed for four lattice-size. Clearly, $\Delta_{F-B}^{P^{max}}$ behaves nonmonotonically with the strength of the onsite interaction. Both the axes are dimensionless.}
    \label{fig:rc_variation}
\end{figure}

In absence of Wannier-Stark term, it was shown that the power from the bosonic battery is always less than that of the fermionic ones in the presence of moderate onsite interactions in the battery Hamiltonian \cite{konar_battery_1}, specifically, when \(|U/J| \lesssim 4\).  Quantitatively, we concentrate on the quantity, 
\begin{equation}
    \Delta^{P^{\max}}_{F-B} = P^{\max}_F - P^{\max}_B,
    \label{eq:deltaPmax}
\end{equation} 
whose negativity guarantees the superiority of the bosonic batteries over the fermionic ones. As indicated in Fig. \ref{fig:roleofrc} (a)(comparing solid lines with the dotted ones), 
as \(r^c_x/U^c_x\) grows, the range of \(|U_x/J_x| \) gets increased in which   \(\Delta_{P^{\max}}^{F-B}\) becomes negative, thereby bosonic batteries outperform the fermionic batteries for a higher range of \(|U_x/J_x|\) provided the initial state is prepared in the ground state of \(H_x\) with \(J_x \neq 0\) and \(U_x\neq 0\). There exists a threshold value of \(r^c_x\), above which \(P^{\max}_{B} \geq P^{\max}_{F}\) for any values of \(|r^c_x/U^c_x|\) (as depicted in Fig. \ref{fig:rc_variation}) irrespective of the filling factor). When the initial state of the battery is prepared in presence of strong hopping strength, the comparable (strong) Wannier-Stark field is required for bosons to outperform the femionic batteries. These results establish that Wannier-Stark field is essential  in order to obtain more power from the bosonic batteries -- we call this upliftment for bosons  as an {\it  activation of power}. 

% \begin{figure}[h]
% \includegraphics[width=0.48\textwidth]
% {J=1,r=0,UVary,r_cVary,U_c=2,BosonsVsFermions.pdf}
% \caption{Difference between maximum power of fermionic and bosonic battery as a function of $\frac{U}{J}$-We have taken J=1, r=0, U=varying from -5 to 5, $U_c=2$. We can see, as $r_c$ increases, bosons tend to perform better.}
% \label{fig:deltaPbvsf}
% \end{figure}

% \begin{figure}
%     \centering
%     \includegraphics[width=0.8\linewidth]
%     {f_b_betaVariation_rc2.pdf}
%     \caption{Variation of $\Delta^{f-b}_{P_{max}}$ with $\frac{U}{J}$; solid line: $\beta = 50$, dashed dot: $\beta = 10$, dotted line: $\beta = 2$}
%     \label{fig:beta_variation}
% \end{figure}

% \begin{figure}[h]
% \includegraphics[width=0.4\textwidth]
% {J=1,r=0,U=3,r_c vary,U_c=2,Pmax,BosonsVsFermions.pdf}
% \caption{Wannier-Stark term in charging Hamiltonian is helpful. We have taken J=1, r=0, U=3, $U_c=2$}
% \label{}
% \end{figure}

%\subsubsection{Battery Hamiltonian (J=1,r=0,U:(-5,5)), Charging (U=2,r=1)}
% In this case, there is a critical value of $r_c$ beyond which no matter what value $U$ takes, bosons always perform better than fermions. for this example, the critical value is close to $\pi$. 

% \subsubsection{Battery Hamiltonian (J=1,r=0,U:3, Charging (U=2,r=(-3,3)}

% In this case, power first decreases(close to $r_c = 1$, there is a minimum), then again increases monotonically. Also incrasing $r_c$ helps in quicker charging and higher storage of power.

% \subsubsection{$i$ vs $i^2$ Wannier-Stark ladder}
% Previously we had taken the Wannier-Stark ladder term as $\sum_{i} r_c \, i \, n_i$. Now we are taking this as $\sum_i r_c \, i^2\,n_i$.
% Here as number of sites increases, critical value of $r_c$ for which bosons are better decreases.

%\subsubsection{What happens if we vary number of particles with fixed number of sites?}
\begin{figure}
    \centering
    \includegraphics[width=\linewidth]{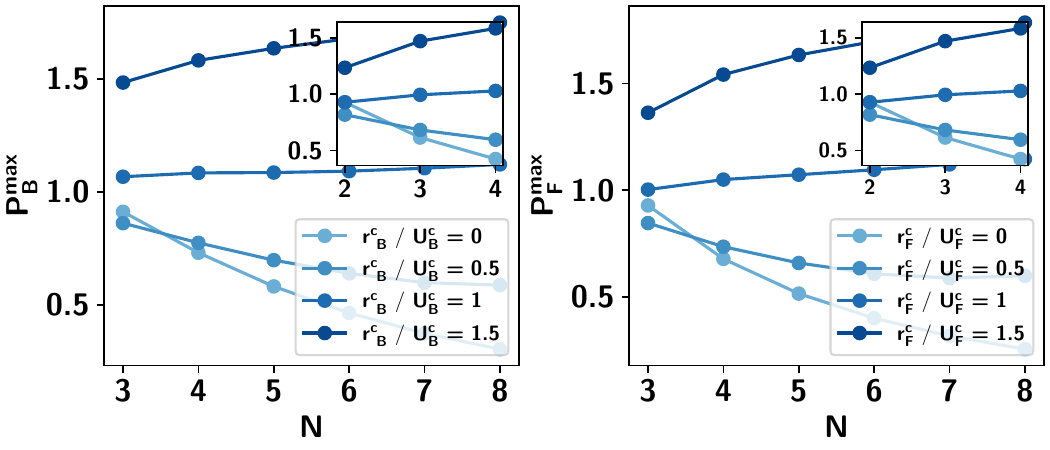}
    \caption{Scaling of \(P_x^{max}\)  (ordinate) (\(x \in \{B\, \text(left),F\, \text(right)\}\))  against system size, \(N\) (abscissa) for a fixed particle number \(n=3\) for bosons and for fermions, \(n_{\uparrow}=2\) and \(n_{\downarrow} =1\). Scaling shows sub-linear behavior although in the presence of Wannier-Stark field, \(P^{\max}_x\) starts increasing with \(N\) instead of decreasing. Other system parameters are \(J_x=1, r_x=0, U_x=3\) in the battery Hamiltonian while different \(r^c_x/U^c_x\) values with \(J^c_x=0\) are chosen to emphasize the beneficial role of Wannier-Stark field. All the axes are dimensionless.}
    \label{fig:scalingfixparticles}
\end{figure}

% \begin{figure}[h!]
%     \centering
%     \includegraphics[width=0.23\textwidth]{fixedParticleBosonPvary.pdf} 
%     % \begin{picture}
%     %     \put(200, 10){\includegraphics[width=0.2\textwidth]{fixed2PsitevaryB.pdf}}
%     % \end{picture}
    
%     \includegraphics[width=0.23\textwidth]{fixedParticleFermionPvary.pdf} 

%     \caption{3 particles, site varies: left boson, right fermion, $J=1,r=0,U=3,J_c=0,U_c=2$}
%     \label{fig:scalingfixparticles}
% \end{figure}

{\it Robustness of activation in power against temperature.} Let us now analyze how this enhancement of power in the bosonic battery is robust against different noisy situations. To probe this, let us prepare the initial state as the canonical equilibrium state of the battery Hamiltonian with nonvanishing temperature, i.e., $\exp (-\beta H_{x})/\tr(\exp(-\beta H_{x}))$ with moderate \(\beta = 1/ k_BT\), (\(k_B\) and \(T\) being the Boltzmann constant and temperature respectively) . In this situation, we observe that activation of power with bosons over fermions can still be  observed and $\Delta_{F-B}^{P^{max}}$ becomes more negative with the increase of \(T\) (see Fig. \ref{fig:rc_variation}). 

\subsection*{Response of Wannier-Stark field on the scaling of power}

Let us discuss some dramatic effects of Wannier-Stark field on storing energy in QB with the increase of filling factor, i.e.,  when the  numbers of particles, \(n=\sum_i\langle \hat{n}_i\rangle\)  are increased for a given lattice-size, and when  \(N\), varies by fixing the number of particles. 

{\it Scaling of power with lattice-size.} It is intriguing to  find how the power scales with the increase of lattice-size, \(N\). In the absence of Wannier-Stark term, we first notice that \(P^{\max}_x\) decreases with the increase of \(N\), irrespective of particle statistics. To explain it, let us first consider the case when there are only two-bosons in the lattice.  Irrespective of the lattice-size, when the battery is in a ground state of \(H_{B}\) with only \(J_B\neq 0 \) and the battery is charged with the Bose-Hubbard Hamiltonian having \(U^c_B \neq 0\), the work stored takes the form 
\begin{eqnarray}
    W_B(t) = \alpha' (N, 2, J_B)  (1-  \cos U^c_B\,t)),
    \label{eq:twoparticles}
\end{eqnarray} 
where \(\alpha'\) is a function of  parameters in the battery Hamiltonian, \(J_B, N\). 
If we introduce Wannier-Stark field in the charging, the expression for work gets modified as mentioned in Eq. (\ref{eq:generalworkwithWS}), where \(\alpha\), \(\beta\) and \(\gamma\) are functions of \(J_B\) and \(N\). It can be shown that \(P^{\max}_B\) with \(U^c_B \neq 0\) and  \(r^c_B =0 \) decreases with the increase of lattice sites while the opposite picture emerges with the increase of \(r^c_B\). Specifically, the power begins to increase as the number of lattice sites increases with the aid of the Wannier-Stark field (see Fig. \ref{fig:scalingfixparticles} for two (inset) and three particles). However, the  scaling of \(P^{\max}_x\) with \(N\) always remains sub-linear, both for bosons and fermions,  when only the nearest-neighbor hopping, onsite and WS field  are present.   %It also raises the possibility that superextensive scaling in this system may be possible as the range of interactions increases. 

%In fact, we discover that this holds true for both bosons and fermions provided one incorporates long-range interactions, i.e., \(H_B\) or \(H_F\) having \(J_B\) and \(J_F\) allowing interactions between arbitrary sites, \(i\) and \(j\). Specifically, we find that \(P^{\max}_x \sim L^{\nu}\) where \(\nu>1\) when the battery Hamiltonian can be represented as and the charging is performed with \(\). \adi{results to be written. bosons!}

{\it Varying particle numbers.} For bosons and fermions, the maximum average power increases with the increase of the number of particles in the presence of onsite interaction without the Wannier-Stark field in the charging. In sharp contrast, although the Wannier-Stark ladder is capable of generating high average maximum power with a single particle,  it converges to a certain power which slowly decreases with the change of number of particles \(n\) in the lattice.  Further, there are parameter regimes in which the power in the absence of \(r^c_x\) (\(x=B,F\)) is  always  lower than that obtained with nonvanishing  \(r^c_x\). Notably, we observe that when \(N =n\),  \(P^{\max}_B > P^{\max}_F\), in the presence of moderate \(r^c_x/U^c_x\) values. 

\begin{figure}
    \centering
    \includegraphics[width=\linewidth]{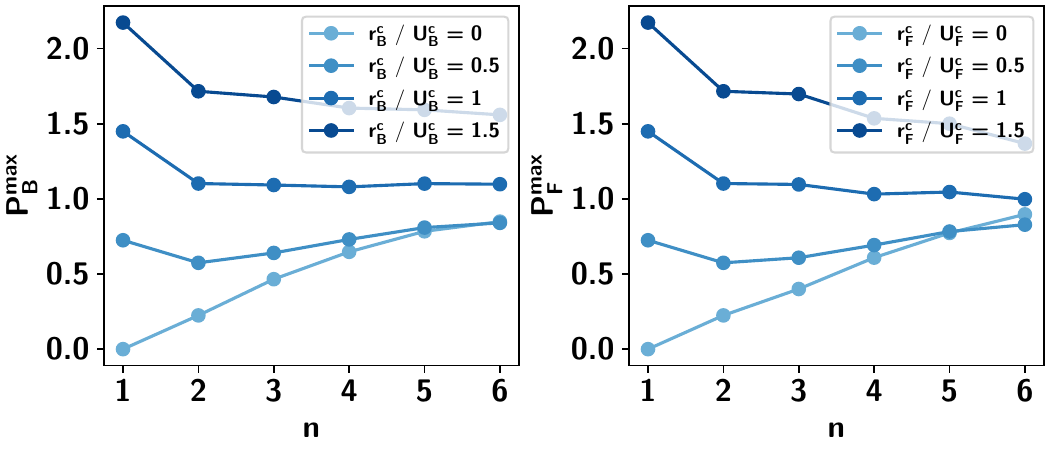}
    \caption{ Scaling of \(P_x^{max}\) (\(x\in\{F (a),B (b)\}\)) (ordinate)  vs particle number, \(n\) (abscissa) for a system-size \(N=6\). It is evident that with the introduction of WS field in the charging, \(P^{\max}_x\) gets enhanced. As shown in Proposition 3, when \(n=1\), \(P^{\max}_x \) is same for bosons and fermions while with the increase of \(n\), we observe that \(P^{\max}_B \geq P^{\max}_F\) in the presence of moderate  \(r^c_x\).  All axes are dimensionless.}
    \label{fig:scaling_fixed_N}
\end{figure}
%Here we have fixed number of sites(N) = 6, and varied number of particles(n) from 1 to 6. Going beyond the half filled case(n = 6) will repeat the same values as maximum power is symmetric with respect to the half filled case. left: boson; right: fermion.

% \begin{figure}[h!]
%   %  \centering
%     % First figure
%    % \begin{subfigure}[b]{}
%         \centering
%         \includegraphics[width=0.23\textwidth]{fixed6siteBosonPvary.pdf}
%           \includegraphics[width=0.23\textwidth]{fixedSiteFermionPvary.pdf}% Replace with your image file
   
% \caption{6 sites: particle varies, left: boson; right: fermion.}
%     \label{fig:particlevaries}
% \end{figure}

% \begin{figure}
%     \centering
%     \includegraphics[width=0.5\linewidth]{fixed6siteBosonPvary.pdf}
%     \caption{6 sites: boson number varies from 1 to 6, \(J =1, U =3, U_c=2\)}
%     \label{fig:enter-label}
% \end{figure}

% \begin{figure}[h]
   
%         \centering
%         \includegraphics[height=1.5in]{fixedParticleBosonPvary.pdf}
%         \caption{3 Bosons, lattice-size is varying from 3 to 8. same as Fig. 4}
% \end{figure}

% \begin{figure}[h]
   
%         \centering
%         \includegraphics[height=1.5in]{bosonParticleVaryr_c=0.pdf}
%         \caption{Bosons: $r_c = 0$, 6 sites, particle number varying from 1 to 6.}
% \end{figure}
    ~ 

\section{Environment-boosted ergotropy in  battery } 
\label{sec:noiseB}

One of the main goal here is to understand the effects of environment upon the quantum battery made out of ultracold atoms which has not been addressed in literature. In this regard, we study the effect of bosonic bath attached at the boundary of the lattice site, given by 
%\begin{equation}
 \(H_{E_i}=\sum_{r=0}^{\omega_c}\eta \hat{a}_r^\dagger \hat{a}_r\), 
%\end{equation}
where \(\hat{a}_r^\dagger\)(\(\hat{a}_r\)) creation (annihilation) operator of mode \(r\) of the oscillator with \([\hat{a}_r,\hat{a}_{r'}^\dagger]=\delta(r-r')\), \(\eta\) is the strength of the interaction and \(\omega_c\) is the cut-off frequency of the bath. We consider the temperature of each bath to be \(T_{E_i}\) (i.e., \(\beta_{E_i}=1/k_BT_{E_i}\)) and the system-bath interactions with the lattice for bosons and fermions  are given respectively as
\begin{eqnarray}
    H_B^{SE_i}&=&\sum_{i}\sum_{r} \hat{n}_i\,\otimes [\hat{a}_r + \hat{a}_r^\dagger],\\
    H_F^{SE_i}&=&\sum_{i}\sum_{r} (\hat{n}_{i\uparrow}+\hat{n}_{i\downarrow})\,\otimes [\hat{a}_r + \hat{a}_r^\dagger],
\end{eqnarray}
where \(\hat{n}_i\) (\(\hat{n}_{i\uparrow},\hat{n}_{i\downarrow}\)) is the number operator at site \(i\) in which the bath is attached. Such a system-environment interaction can be considered as a generalized dephasing model in the ultracold atomic setup. As the system is connected to  the thermal bath, the evolution of the system is governed by the Gorini–Kossakowski–Sudarshan–Lindblad  (GKSL) master equation \cite{breuer2002, rivas2012},
\begin{equation}
\dot{\hat{\rho}}=-\text{i}[H^{c},\hat{\rho}]+\mathcal{L}(\hat{\rho}),
\label{eq:qme}
\end{equation}
where \(H^{c}\) is the charging Hamiltonian and \(\mathcal{L}(\hat{\rho})=\sum_i L_i\) is the dissipation part due to the environment. Here each \(L_i\) can be represented as
\begin{eqnarray}
L_i(\hat{\rho}) &=&\nonumber \sum_{\omega}\gamma_i(\omega) \left[A_i(\omega)\hat{\rho} A_i^\dagger(\omega)-\frac{1}{2}\left\{A_i^\dagger(\omega)A_i(\omega),\hat{\rho} \right\}\right],\\
\label{lindblad_global}
\end{eqnarray}
where \(\gamma_i(\omega)\) is the time-dependent dephasing transition rate of the \(i\)th bath quantifying the environment correlation, 
  
\begin{eqnarray}
\gamma_i(\omega) &=& f_i(\omega)[1+\kappa_i(\omega)],\text{ for }\omega\geq 0,\nonumber \\
\gamma_i(\omega) &=& f_i(|\omega|)\kappa_i(|\omega|), \text{ for }\omega<0,
\end{eqnarray}
with $f_i(\omega)=\eta_i\omega\exp(-\omega/\omega_c)$, $\eta_i$ being a constant for the $i$th bath and under Markovian approximation, $\max\{\eta_i\}\ll 1$. On the other hand, $\kappa_i(\omega)=\left[\exp(\beta_i\omega)-1\right]^{-1}$,
%\begin{equation}
%\end{equation}
% \begin{equation}
%     \gamma_i(\omega,t)= J(\omega)\coth(\frac{\omega}{2k_BT_{E_i}})\frac{\sin \omega t}{\omega},
% \end{equation}
with \(\omega\) being the transition frequency of the bath. Here \(A_i(\omega)\) is the Lindblad operator for the bath attached to the  site \(i\) corresponding to the transition  energy $\omega$ among the energy levels of the battery, \(H_x\), i.e.,  
\begin{eqnarray}
\text{e}^{\text{i}H_x t}n_i\text{e}^{-\text{i}H_xt}=2\sum_{\omega}A_i(\omega)\text{e}^{-\text{i}\omega t}.
\end{eqnarray}
Due to the detrimental effect of environment, the work stored in the battery may not be the same as the extractable work as some portion of the work is dissipated as heat while obeying the first law of thermodynamics. Therefore, the maximum extractable work in a given state, \(\hat{\rho}\), called as {\it ergotropy} is  defined as \cite{Alicki, battery_rmp_review}
\begin{equation}
    \mathcal{E}(t)=W_x(t)-\underset{U}{\max}\Tr[H_x U\hat{\rho} U^\dagger],
\end{equation}
where the maximization in the definition can be computed by writing both \(H_x\) and \(\hat{\rho}\) in the eigenbasis of \(H_x\).

% where \(T_{E_i}\) is the temperature of the bath, \(k_B\) is the Boltzmann constant and \(J(\omega)\) is the  Ohmic-like spectral densities of the environment, given as
% \begin{equation}
%     J(\omega)=\frac{\omega^s}{\omega_c^{1-s}}e^{-\omega/\omega_c}
% \end{equation}
% where \(\omega_c\) is the cut-off frequency of the environment and \(s\) is the parameter quantifies sub-ohomic (\(s<1\)), ohomic (\(s=1\)) and super-ohomic (\(s>1\)) reservoirs and \(A_i(\omega)\) is the Lindblad operator for the bath attached with the \(i\)th site for the transition amounting energy $\omega$ among the energy levels of the battery,\(H_B\), which is given as  
% \begin{eqnarray}
% \text{e}^{\text{i}H_{B}t}n_i\text{e}^{-\text{i}H_{B}t}=2\sum_{\omega}A_i(\omega)\text{e}^{-\text{i}\omega t},
% \end{eqnarray}
% where \(H_B\) is the battery Hamiltonian. Here we like to point out at \(T_{E_i}=0\) all the frequency dependent part can be integrated out and given as
% \begin{equation}
%     \gamma(T=0,t)=\omega_c[1+(\omega_ct)^2]^{-s/2}\Gamma[s]\sin[s\tan^{-1}\omega_ct]
% \end{equation}
% and the the evolution of the system obey the master equation,
% \begin{equation}
%     \dot\rho=-\text{i}[H_{C},\rho]+\gamma(T=0,t)\sum_i(n_i\rho n_i-\frac{1}{2}\{n_i^2,\rho\}),
% \end{equation}
% which is a generalized dephasing operation, captures the non-markovian effect on ultra-cold atom quantum battery by varying the ohomic parameter \(s\).

\subsection{Effect of ergotropy in presence of Wannier-Stark interaction}

As highlighted in the previous sections, the parameters of both the QB and charging Hamiltonian are vital for achieving higher maximum power. Therefore, it is important to closely examine how these system parameters influence the performance of QB which is now attached to the environment. A deeper understanding of their role can significantly contribute in optimizing the QB’s efficiency and overall performance, leading to better outcomes in the open quantum scenario,
%. Hence, we study the effect of system parameters on this set up
both in charging as well as discharging of the QB.

\begin{figure}
    \centering
    \includegraphics[width=\linewidth]{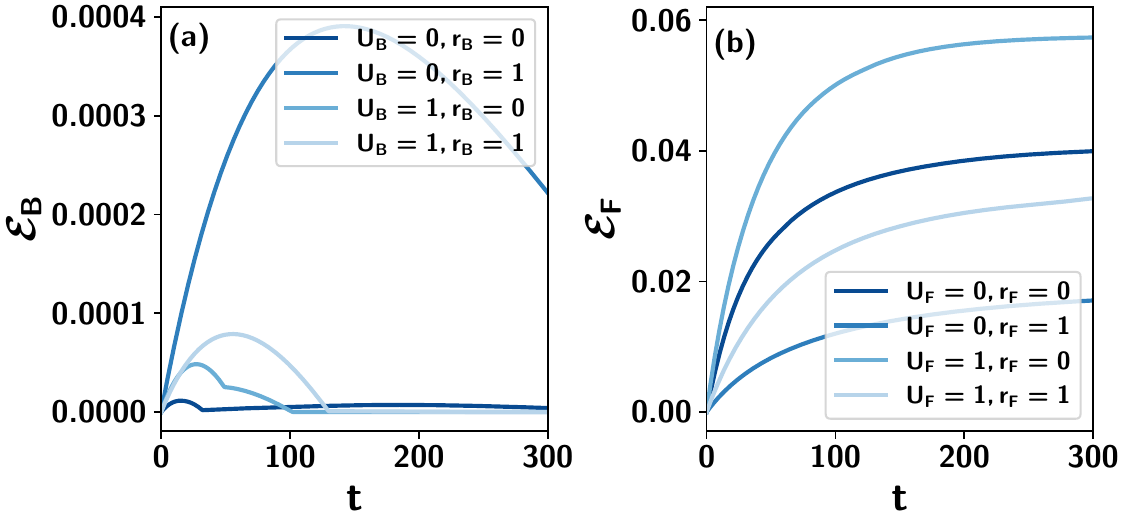}
    \caption{{Charging process.} Ergotropy (ordinate) with time (abscissa) for bosons with  with \(N=n=4\) (a) and fermions (b) \(n_{\uparrow} = n_{\downarrow} =2\) with four sites --{\it Environment-assisted ergotropy}.  Here the initial state is in the ground state of the QB which contains either \(J_x\neq 0\) or \(J_x\neq 0\) and \(r_x\neq 0\) or \(J_x\neq 0\), \(U_x\neq 0\) or all of them are nonvanishing while no unitary driving is present, i.e., $J^c_F = r^c_F = U^c_F = 0$. The steady state ergotropy increases when onsite interaction present in the QB for fermions while the same occurs for bosons when Wannier-Stark field is present in the battery. Other parameters of the setting are \(\forall i, T_{E_1}=T_{E_L}=1\) and \(\eta_i=10^{-2}\).
    %(Inset of (b)) It depicts the increase of the ergotropy when the temperature in both the baths or one of the baths is increased.  
    All the axes are dimensionless.}

   % We have taken 2 fermions in 2 sites: No charger is connected to the battery. Charging is happening only by assistance of environment, $J^c_F = r^c_F = U^c_F = 0$}
    \label{fig:fermi_open_no_charge}
\end{figure}

% \begin{figure}
%     \centering
    %\includegraphics[width=\linewidth]{charging.pdf}
    % \caption{Ergotropy (ordinate) is plotted with time (abscissa). {\it Environment assisted charging} when the initial state in ground state of QB. Rate of increment of ergotropy depends on the battery parameter which is denoted by the different color schemes. Steady state ergotropy increases while onsite interaction present in the QB. {\it Environment assisted charging} is very small in BH model while it is higher in FH model. Other system parameters are \(J_x=1\) and \(\beta_1=\beta_2=1\). All the axes are dimensionless.}
    % \label{fig:charging}
%\end{figure}

\subsubsection{Environment-induced ergotropy for fermions} 

Let us first demonstrate that  the battery can provide the maximal extractable work with the help of system-environment interactions -- we call this  concept as {\it environment-induced ergotropy}.  To achieve this goal, the initial state is prepared in the ground state of  the battery Hamiltonian, \(H_F=-J_F\sum_{\langle ij\rangle} \hat{c}_{i\sigma}^\dagger \hat{c}_{j\sigma} + \text{h.c.}\). It is then subjected  to the influence of the environment as described before, without any external driving forces.  In this case, we observe that the ergotropy of the battery gradually increases over time as a result of environmental interactions. Eventually, the ergotropy saturates to a fixed value at certain time. This steady-state condition represents a form of charging where the environment plays a crucial role in facilitating the transfer of energy into the battery, known as {\it environment-induced ergotropy} of the quantum battery, highlighting the significant contribution of the environment to the energy storage process. 

Under different initial conditions, i.e.,  different choices of battery parameters, \(J_F, U_F\) and \(r_F\) as well the temperature of the bath, \(T_{E_i}\),  the qualitative behavior of the ergotropy remains same,  although the values change with the variation of these parameters as illustrated in Fig. \ref{fig:fermi_open_no_charge} (b).  Based on the numerical simulations, some crucial observations emerge on the parameter-dependence  -- (1) When  temperatures of both the bath are same, the ergotopy increases with the increase of the bath-temperature;  (2) Interestingly, by fixing temperature of a bath, say \(T_{E_1}\), if one decreases (increases) the temperature of another bath to \(T_{E_2}\), i.e., \(T_{E_2} < T_{E_1}\) (\(T_{E_2} > T_{E_1}\)), the extractable work decreases (increases); (3) The saturated ergotropy value increases when both hopping and onsite interactions are present in the battery compared to that obtained from the battery having either hopping  or  hopping and  the WS field.  Further,  {\it environment-assisted ergotropy} is intrinsic to the Fermi-Hubbard model which we cannot find  in the Bose-Hubbard model in the steady state regime. However, in the transient regime, bosonic battery with the help of WS field can also generate nonvanishing ergotropy although the values is very small compared to that obtained from fermions (see Fig. \ref{fig:fermi_open_no_charge} (a)).
%\, thereby  demonstrating the usefulness of the FH battery over the BH battery in the presence of environment. 
However, when the charger is incorporated  to the battery, both BH and FH batteries can generate high values of ergotropy in the steady state.

\subsubsection{Charging and discharging in presence of environment} 

\begin{figure}
    \centering
    \includegraphics[width=\linewidth]{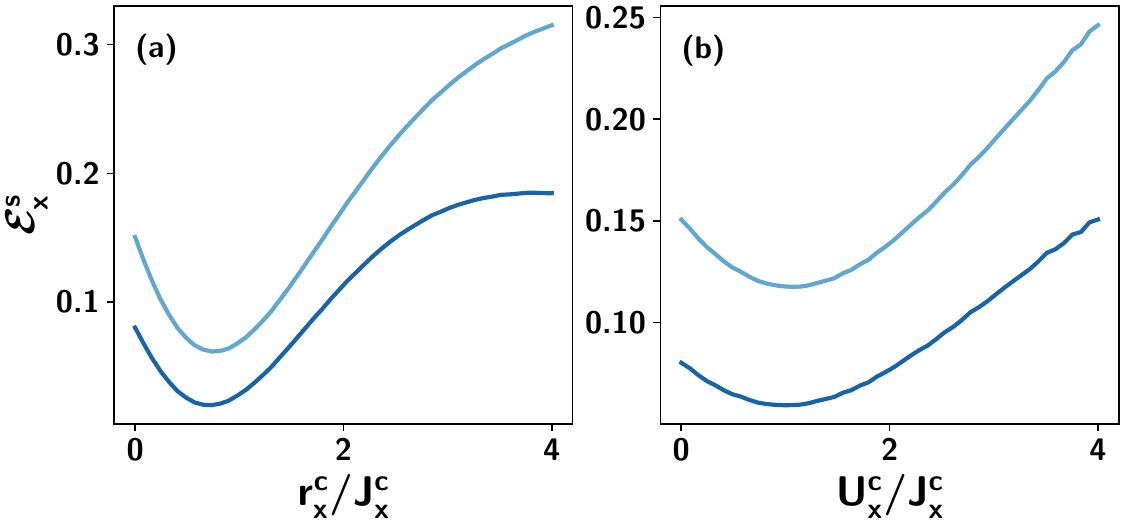}
    \caption{{\it Nonmonotonicity with charging strength.}The ergotropy in the steady state (vertical axis) with respect to \(r^c_x/J^c_x\) (a) and \(U^c_x/J^c_x\) (b) in presence of environment.  The initial state is prepared as the ground state of the battery Hamiltonian with $J_x = r_x = U_x = 1, N = 2, n = 2$ for bosons and $n_\uparrow = n_\downarrow = 1$ for fermions.  Dark (black) line specifies the BH model while the light (gray) line  corresponds to the FH model. All other parameters are same as in Fig. 
    \ref{fig:fermi_open_no_charge}. Both the axis are dimensionless. }
    \label{fig:bose}
\end{figure}

It is expected that the environment has a detrimental impact on the charging process of the quantum battery, reducing the amount of extractable work. Therefore, it is crucial to examine the charging Hamiltonian, which governs the unitary evolution of the system. Specifically, the interplay between the unitary dynamics and environmental dissipation can influence the steady-state ergotropy, with this effect being dependent on the strength of the charging parameters. To explore this, we analyze the behavior of ergotropy during the charging process by varying key charging parameters, namely  \(J^c_x\), \(U^c_x\) and \(r^c_x\) (\(x=B, F\)). Our results indicate that, regardless of whether the battery is described by the BH or FH model, increasing the onsite charging strength or WS field strength can lead to an increase in the ergotropy after a certain threshold value. It indicates that this increase is not monotonic with \(U^c_x\) or \(r^c_x\), as demonstrated in Fig. \ref{fig:bose}. This observation suggests that, while environmental assistance in the charging process is important, the parameters of the charging Hamiltonian itself — specifically the strength of the onsite charging interaction — play a critical role in determining the efficiency of the charging of QB in the presence of environmental influences. Thus, optimizing these charging parameters is essential for enhancing the performance of the quantum battery and mitigating the detrimental effects of the environment.

% Environment plays a detrimental effect on charging of the QB which tells us after charging extractable work become less, hence, it is important to look at the charging Hamiltonian which constitute the unitary evolution. More precisely, the competition between unitary and environmental dissipation can enhance the steady state ergotropy which depends upon the charging parameters strength. Hence, we observe the effect of ergotropy while charging of QB by varying charging parameters, \(J^c_F\), \(U^c_F\) and \(r^c_F\). We find that irrespective of BH and FH battery by varying onsite charging strength (\(U^c_F\)) can increase ergotropy while such increment non monotonic as shown in Fig. \ref{}. One can find out that apart from environment assisted charging, strength of the parameters of the charging Hamiltonian plays essential role in charging of QB in presence of environment. 

\begin{figure}
    \centering
    \includegraphics[width=\linewidth]{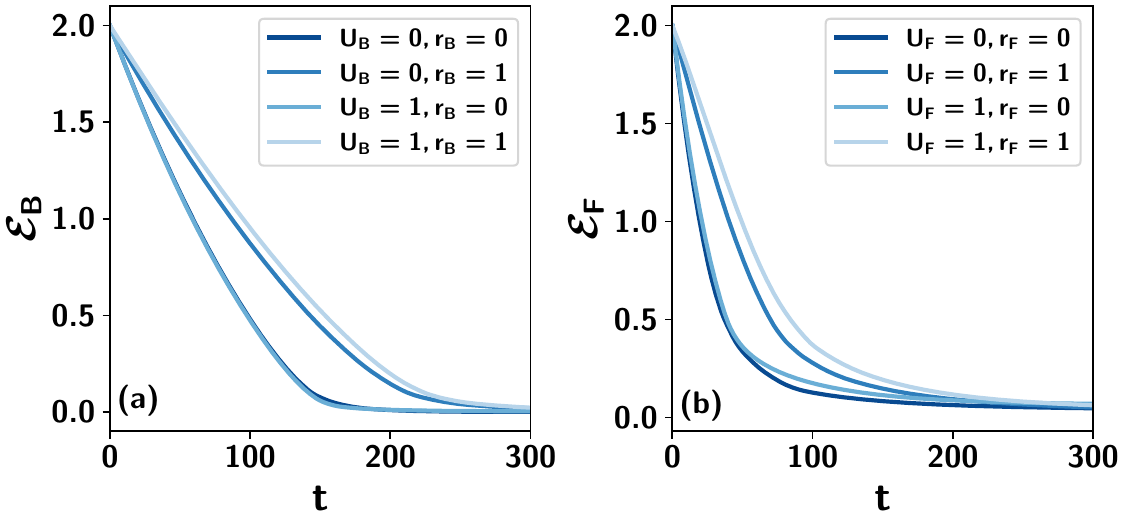}
    \caption{{\it Discharging process.} Ergotropy (vertical axis)  with time (abscissa)  when the initial state is in the maximally charged state having maximum ergotropy. (a) bosons and (b) fermions. Rate of decrement of the ergotropy depends on the battery parameters as shown in legends. Slow rate in ergotropy decrement is observed when both WS and onsite interactions  are present in the battery Hamiltonian.  Other system parameters are \(J_x=1\), \(\beta_1=\beta_2=1\), \(N=4\), \(n=4\) for bosons and  \(n_{\uparrow} = n_{\downarrow} =2\). All the axes are dimensionless.}
    \label{fig:discharging}
\end{figure}

{\it Discharging in presence of environment.} In a battery, especially when it is in contact with the bath, charging and discharging can be different, i.e., the stored energy may not be extractable completely during discharging of the battery. Hence, it is intriguing to determine how the quantum battery discharges, specifically focusing on the decay in ergotropy  when the battery is fully charged. The rate at which the ergotropy diminishes and its detrimental effects provide critical insights into the battery's robustness against environmental factors.
Our findings indicate that the ergotropy decays more rapidly in the absence of Wannier-Stark field and onsite interactions compared to the cases when one of them  or both of them is present  (see Fig. \ref{fig:discharging}). In particular, we observe that when either the WS field or the onsite interaction and WS field along with the hopping is incorporated in the battery,  the ergotropy decreases slowly as compared to the cases when \(J_x\) or \(J_x, U_x \neq 0\) for bosons and fermions. This suggests that these factors contribute positively to the QB's performance. 

{\it Discharging in presence of a charger.} Additionally, when unitary driving is introduced, we observe that the ergotropy decays but eventually reaches a steady-state value, indicating a balance between the unitary driving forces and dissipation. This saturation behavior highlights the competition between the unitary dynamics (first term in Eq. (\ref{eq:qme})) and the dissipative processes (second term in Eq. (\ref{eq:qme})), which influence the overall energy retention in the system. These results emphasize the importance of specific QB parameters in mitigating environmental effects and improving performance, while also showcasing the interplay between dissipation and unitary dynamics in determining the long-term behavior of the quantum battery. 

\section{Conclusion}
\label{sec:conclu}

The moderate strength of the Wannier-Stark (WS) potential in the cold atomic setup in an optical lattice induces changes such as the system's regular or irregular spectrum, providing a platform to investigate atom-atom interactions in the laboratory.   We employed this static electric field to explore the possibility of increasing the energy storage capacity of quantum batteries (QB). Indeed, we found that the Wannier Stark field in the charging Hamiltonian has an influence on generating power in the battery both in bosonic and femionic batteries, when its initial state is prepared in the ground or thermal state of the Hubbard models with hopping and onsite interactions. However, we observed that at a certain threshold value of the WS field, the power produced by QB for bosons turns out to be higher than that of the fermions, in contrast to quantum batteries without a WS ladder -- {\it activation of power via WS field for bosons over fermions}. The bosonic batteries outperform more than the fermionic ones when the initial state is in a thermal equilibrium state with a  moderately high temperature. We also found that the power 
of the QB can be enhanced by using WS field when either the number of particles grows or  when the  lattice-size increases. 

When the battery setup remains affected by the environment, it adversely impacts the performance of the battery. Nevertheless, we demonstrated that when the edge of the batteries are connected to a thermal bath with a temperature ranging from moderate to high, it is feasible to extract work from the fermionic batteries in the steady state without requiring to undergo a charging procedure -- we call it as {\it environment-assisted ergotropy}. Such a phenomena is observed for bosons only in the transient regime in the presence of WS field. Further, ergotropy does not always rise with an increase in onsite interaction strength (WS field strength) when both unitary driving and dissipative dynamics are present; rather, it grows with an increase in any of them above a critical strength.  Moreover,  we also exhibited that starting from the highest excited state, the rate at which the ergotropy decreases  in the presence of the WS field is slower   than in its absence for both  bosonic and fermionic batteries.  It is fascinating to study  how the environment's beneficial effects might be leveraged to construct further quantum devices on this cold atomic substrate.

\acknowledgments

We  acknowledge the use of the cluster computing facility at the Harish-Chandra Research Institute. This research was supported in part by the INFOSYS scholarship for senior students. LGCL received funds from project DYNAMITE QUANTERA2-00056 funded by the Ministry of University and Research through the ERANET COFUND QuantERA II – 2021 call and co-funded by the European Union (H2020, GA No 101017733).  Funded by the European Union. Views and opinions expressed are however those of the author(s) only and do not necessarily reflect those of the European Union or the European Commission. Neither the European Union nor the granting authority can be held responsible for them. This work was supported by the Provincia Autonoma di Trento, and Q@TN, the joint lab between University of Trento, FBK—Fondazione Bruno Kessler, INFN—National Institute for Nuclear Physics, and CNR—National Research Council.

\bibliography{reference}

\end{document}